\documentclass[aps,prl,showpacs,floatfix,twocolumn,superscriptaddress]{revtex4-1}
\usepackage{graphicx}
\usepackage{bm,amsmath}
\usepackage{dcolumn}
\usepackage{amsfonts,amssymb}
\usepackage{chemfig}
\usepackage{diagbox}

\begin{document}

\title{Consistent scaling exponents at the deconfined quantum-critical point}

\author{Anders W. Sandvik}
\email{sandvik@bu.edu}
\affiliation{Department of Physics, Boston University, 590 Commonwealth Avenue, Boston, Massachusetts 02215, USA}
\affiliation{Beijing National Laboratory for Condensed Matter Physics and Institute of Physics, Chinese Academy of Sciences, Beijing 100190, China}

\author{Bowen Zhao}
\affiliation{Department of Physics, Boston University, 590 Commonwealth Avenue, Boston, Massachusetts 02215, USA}

\date{July 7, 2020}

\begin{abstract}
We report a quantum Monte Carlo study of the phase transition between antiferromagnetic and valence-bond solid ground states in the 
square-lattice $S=1/2$ $J$-$Q$ model. The critical correlation function of the $Q$ terms gives a scaling dimension corresponding to
the value $\nu = 0.455 \pm 0.002$ of the correlation-length exponent. This value agrees with previous (less precise) results from conventional
methods, e.g., finite-size scaling of the near-critical order parameters. We also study the $Q$-derivatives of the Binder cumulants
of the order parameters for $L^2$ lattices with $L$ up to $448$. The slope grows as $L^{1/\nu}$ with a value of $\nu$
consistent with the scaling dimension of the $Q$ term. There are no indications of runaway flow to a first-order phase transition.
The mutually consistent estimates of $\nu$ provide compelling support for a continuous deconfined quantum-critical point.
\end{abstract}

\maketitle

Among the many proposed exotic quantum states and quantum phase transitions beyond the Landau-Ginzburg-Wilson (LGW) paradigm in two dimensions
\cite{Sachdev08,Savary17,Wen19}, the deconfined quantum crititical point (DQCP) \cite{Senthil04} is special because it has concrete lattice
realizations in sign-free ``designer models'' accessible to quantum Monte Carlo (QMC) simulations \cite{Kaul13}. Indeed, the first hints of an
LGW-forbidden continuous transition between antiferromagnetic (AFM) and spontaneously dimerized valence-bond solid (VBS) ground states
came from QMC simulations \cite{Sandvik02}, and following the DQCP concept 
(which builds on previous works on VBS phases and topological defects \cite{Haldane88,Chakravarty89,Read89,Read90,Dagotto89,Murthy90,Motrunich04}) 
numerous additional studies have been reported. The most compelling results for DQCP physics have been obtained with $J$-$Q$ models
\cite{Sandvik07,Melko08,Jiang08,Lou09,Sandvik10a,Kaul11,Sandvik12,Harada13,Chen13,Block13,Pujari15,Suwa16,Shao16,Ma18}, which are Heisenberg 
antiferromagnets in which the exchange of strength $J$ is supplemented by multi-spin couplings of strength $Q$ that induce 
singlet correlations and eventually cause spontaneous dimerization. The space-time loop structure employed in QMC simulations
\cite{Evertz03,Sandvik99,Sandvik10b} can also be used to formulate analogous classical three-dimensional loop models, 
which exhibit behaviors very similar to the $J$-$Q$ models \cite{Nahum15a,Nahum15b}.

Even though very large lattices have been studied, with linear size $L$ up to $256$ for the $J$-$Q$ model \cite{Sandvik10a,Shao16}
and twice as large for the loop model \cite{Nahum15b}, it has not yet been possible to draw definite conclusions on the nature of 
the AFM--VBS transition. While no explicit signs of a first-order transition have been detected in the best DQCP candidate models
(in contrast to intriguing discontinuous transitions with emergent symmetry in related models \cite{Zhao19,Serna19,Takahashi20}),
some observables exhibit scaling behaviors incompatible with conventional quantum criticality. Such behaviors have been interpreted as runaway
flows toward what would eventually become a first-order transition on lattices even larger than those studied so far \cite{Jiang08,Kuklov08,Chen13}. 
Another proposal is that the DQCP is even more exotic than initially anticipated, with novel relationships between critical exponents originating 
from the presence of two divergent langth scales \cite{Shao16}---in addition to the standard correlation length $\xi$, there is a larger scale $\xi'$ 
associated with a ``dangerously irrelevant'' perturbation and emergent U(1) symmetry of the near-critical VBS fluctuations in the DQCP 
scenario \cite{Senthil04,Levin04}. The weak first-order scenario has attracted attention in the context of non-unitary conformal 
field theories (CFTs), which have critical points in the complex plane \cite{Wang17,MaHe19,Ma19b,Nahum19}. In this scenario, the AFM--VBS transition is a 
``walking'' first-order transition \cite{Gorbenko18a,Gorbenko18b} where the renormalization-group flow (which is manifested also in finite-size scaling) 
is affected by the inaccessible nearby critical point and only slowly ``walks'' to a first-order instability.

In support of the weak first-order scenario, the $J$-$Q$ and loop models are often invoked as supporting evidence, 
though there are no concrete predictions that have been compared with the numerical results. In the absence of any quantitative tests or clear 
signs of discontinuities or coexistence state in the lattice models, the walking scenario should not be accepted as the final word on 
the fate of the DQCP. Here we will show that the cited \cite{MaHe19,Ma19b,Nahum19} large scaling corrections affecting estimates of the
critical correlation exponent $\nu$ \cite{Harada13,Shao16,Nahum15b} are not precursors to a first-order transition. We reach this 
conclusion by extracting $\nu$ from the scaling dimension of the relevant field of the model. The corresponding correlation function exhibits 
only small scaling corrections and delivers an exponent compatible with results based on Binder cumulants; $\nu =0.455(2)$. Given the
well behaved estimators of $\nu$, a continuous transition is the most likely scenario.

To set the stage, we briefly summarize some standard facts on critical scaling. Consider a Hamiltonian $H_c$ tuned to a 
quantum critical point to which a perturbation is added that maintains all the symmetries of $H_c$;
\begin{equation}
H = H_c + \delta \sum_\mathbf{r} D(\mathbf{r}),
\label{HD}
\end{equation}
where $\mathbf{r}$ denotes the lattice coordinates and $D(\mathbf{r})$ are local operators. Normally $H$ is written in a form with
some tunable parameter $g$ such that, for some critical value $g=g_c$, $H(g_c)=H_c$ and $\delta=g-g_c$.
We assume that the system develops long-range order when $\delta>0$, with an order parameter $m(\mathbf{r})$ such that 
$\langle m\rangle=\langle m(\mathbf{r})\rangle \propto \delta^\beta$ for small $\delta>0$ and $m=0$ for $\delta<0$.  The critical exponent 
$\beta$ depends on the universality class of $H_c$ in the thermodynamic limit. On either side of the phase 
transition, the exponential decay of the correlation function $C_m(\mathbf{r})=\langle m(0)m(\mathbf{r})\rangle-\langle m\rangle^2$ defines the 
divergent correlation length, $\xi \propto |\delta|^{-\nu}$. At $\delta=0$, the correlation function takes the critical form 
$C_m(r) \propto r^{-2\Delta_m}$, where $\Delta_m=\beta/\nu$ is the scaling dimension of the operator $m$.

In QMC calculations $\nu$ is typically extracted using finite-size scaling of some dimensionless quantity, such as the Binder 
ratio $R = \langle M^4\rangle/\langle M^2\rangle^2$, where $M=\sum_\mathbf{r}  m(\mathbf{r})$. Neglecting scaling corrections, in the 
neighborhood of the critical point we have $R(\delta,L)=R(\delta L^{1/\nu})$, by which $\nu$ (and the critical point $g_c$ if it is not known) 
can be obtained from data for different values of $\delta$ and and $L$. A less common method is to use the relation
$1/\nu = d-\Delta_D$, where $d$ is the space-time dimensionality (here $d=3$) and $\Delta_D$ is the scaling dimension of the perturbing
operator $D$ in Eq.~(\ref{HD}). The scaling dimension can be obtained from the power-law decay $C_D(r) \propto r^{-2\Delta_D}$ of the correlation
function $C_D(\mathbf{r})=\langle D(0)D(\mathbf{r})\rangle-\langle D\rangle ^2$ at $g_c$.

It is not clear to us why $\nu$ is not commonly extracted from $C_D(\mathbf{r})$, but there are two potential drawbacks: 
(i) Often $\Delta_D$ is rather large, e.g., in the case of the O(3) universality class (of which we will show an example below)
$\Delta_D \approx 1.6$, so that the correlation function decays rapidly and is difficult to compute precisely (with small
relative statistical errors) at large $r$. (ii) The operator $D$ is often off-diagonal and may appear to be technically difficult to compute.
However, although the latter issue is absent in simulations of classical systems, the scaling dimension $\Delta_D$ is still normally not computed. 

Here we will take advantage of the fact that existing estimates of $\nu$ at the DQCP ($\nu \approx 0.45$ in both the $J$-$Q$ \cite{Shao16} and 
loop \cite{Nahum15b} models) correspond to a rather small value of the scaling dimension, $\Delta_D \approx 0.8$, and therefore it may be possible 
to compute it reliably in this case (as was done recently for the transverse-field Ising chain, where, in the notation used here, $\Delta_D=1$ 
\cite{Patil17}). Furthermore, we point out that off-diagonal correlation functions of operators that are terms of the Hamiltonian have very simple 
estimators within the Stochastic Series Exapansion (SSE) QMC method \cite{Sandvik91,Sandvik92,Sandvik99,Sandvik10b}. The quantum fluctuations are 
here represented by a string of length $n$ of terms $H_i$ of $H$, with mean length $\langle n\rangle = |\langle H\rangle|/T$, where $T$ is the 
temperature. A connected correlation function of any two terms is given by \cite{Sandvik92}
\begin{eqnarray}
C_{\rm ab} & \equiv &\langle H_aH_b\rangle  - \langle H_a\rangle \langle H_b\rangle \nonumber \\
           & = & T^2\bigl (\langle n_{ab}(n-1)\rangle - \langle n_a\rangle \langle n_b\rangle \bigr),
\end{eqnarray}
where $n_a$ is the number of operators $H_a$ in the string and $n_{ab}$ is the number of times that $H_a$ and $H_b$ appear adjacent to each other.
This expression can be easily applied to all location pairs $(a,b)$ in a single scan of the operator string, and translational invariance 
can be exploited at no additional cost to improve the statistics.

\begin{figure}[t]
\centering
\includegraphics[width=65mm]{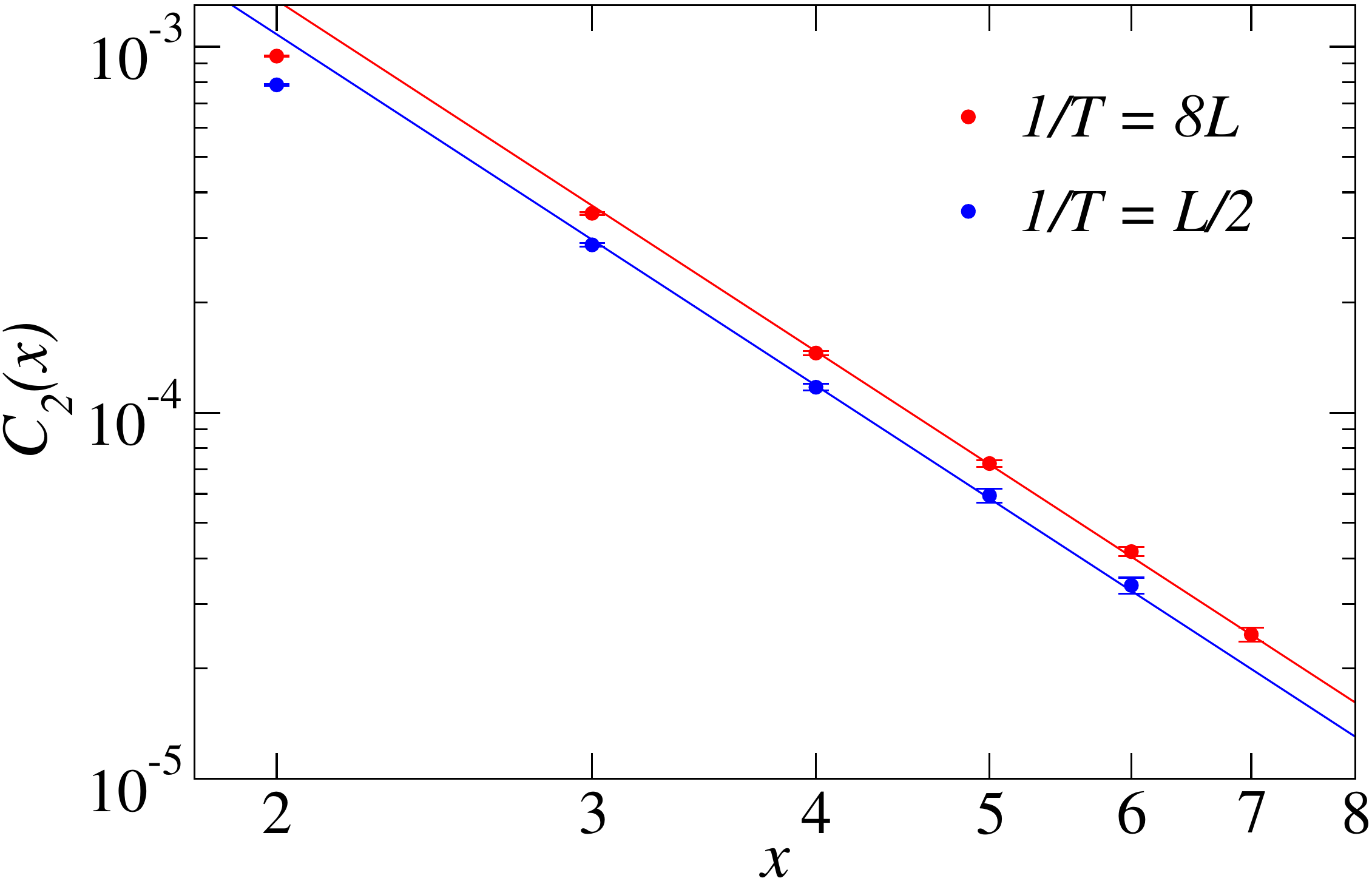}
\caption{Dimer correlation function, Eq.~(\ref{cdbilayer}), in the critical bilayer at separation 
$\mathbf{r}=(x,0)$ with $x=L/2-1$. Results are shown for two different values of $LT$. The lines have slope 
$-2\Delta_2=-3.188$, corresponding to the O(3) value of $\nu$.}
\label{fig:bilayer}
\end{figure}

As a demonstration of the method, we first consider the $S=1/2$ bilayer Heisenberg Hamiltonian
\begin{equation}
H = J_1 \sum_{a=1,2}\sum_{\langle ij\rangle } \mathbf{S}_{a,i} \cdot \mathbf{S}_{a,j}
+ J_2 \sum_{i=1}^N \mathbf{S}_{1,i} \cdot \mathbf{S}_{2,i},
\end{equation}
where $\langle ij\rangle$ denotes nearest-neighbors on a square periodic lattice with $N=L^2$ sites and $a$ is the layer index.
This system has an AFM ground state for $g \equiv J_2/J_1 < g_c$ and is a quantum paramagnet dominated by inter-layer singlet formation for $g>g_c$.
The O(3) quantum phase transition has been investigated in many previous works. Here we take $g_c =2.52205$ for the critical point \cite{Wang06,Sen15}
and study a correlation funtion corresponding to the perturbation $D$ in Eq.~(\ref{HD}). Since both the 
$J_1$ and $J_2$ interactions drive the system away from the critical point, we can study correlations between either type of terms (i.e., they 
have the same scaling dimension). We use the $J_2$ terms, which form a simple square lattice, and define
\begin{equation}
  C_2(\mathbf{r}_{ij}) \equiv \langle (\mathbf{S}_{1,i} \cdot \mathbf{S}_{2,i})(\mathbf{S}_{1,j} \cdot \mathbf{S}_{2,j})\rangle
  - \langle \mathbf{S}_{1,i} \cdot \mathbf{S}_{2,i}\rangle^2,
\label{cdbilayer}
\end{equation}
where $\mathbf{r}_{ij}$ denotes the separation of the sites $i$ and $j$. 

Investigating the decay of the correlations, we can either study large lattices and focus on $r \ll L$ to eliminate
finite-size effects or take $r$ of order $L$ and study the size dependence. Here we opt for
the latter method with $\mathbf{r} = (L/2-1,0)$, for which there are more equivalent points for averaging than for the
high-symmetry points $(L/2,0)$ and $(L/2,L/2)$. For the expected O(3) universality class in 2+1 dimensions $\nu \approx 0.711$
\cite{Compostrini02}, corresponding to a scaling dimension $\Delta_2 \approx 1.594$ of the $J_2$ interaction. As shown in Fig.~\ref{fig:bilayer}, 
because of the rapid decay we can access only rather modest distances, but the results still show a remarkably good agreement
with the expected form $C_2(r) \propto r^{-2\Delta_2}$ starting from $r = 4$ ($L=10$). In the SSE simulations we have used $T=c/L$
(in units with $J_1=1$), reflecting the emergent Lorentz invariance of the system (i.e., the dynamic exponent $z=1$), with two
different proportionality factors; $c=2$ and $c=1/8$. Apart from the different amplitudes of the correlations, both data sets exhibit the same decay. 

Turning now to the $J$-$Q$ model,
we express the AFM Heisenberg interaction as a singlet projector, $-P_{ij}$, on $S=1/2$ spins;  
$P_{ij}=1/4 - \mathbf{S}_{i} \cdot \mathbf{S}_{j}$. To simplify the notation, we use a bond index $b$ to implicitly refer to two nearest-neighbor 
spins $\langle i,j\rangle_b$; $P_b \equiv P_{ij}$. We also use an index $p$ to refer to a $2\times 2$ plaquette with sites in the arrangement
$(\begin{smallmatrix}i & j \\ k & l\end{smallmatrix})_p$ and define $Q_p \equiv P_{ij}P_{kl} + P_{ik}P_{jl}$.
With these definitons the $J$-$Q$ Hamiltonian is \cite{Sandvik07}
\begin{equation}
H = -J \sum_{b} P_{b} - Q \sum_{p} Q_{p}.
\label{jqham}  
\end{equation}
We define the coupling ratio $g \equiv J/Q$ and
use the SSE method to compute the $z$ component of the staggered magnetization (the AFM order parameter)
\begin{equation}
m_z = \frac{1}{N} \sum_\mathbf{r} S^z_\mathbf{r} (-1)^{\mathbf{r}_x+\mathbf{r}_y}, 
\end{equation}
and the two-component dimer (VBS) order parameter, also defined with the $z$ spin components,
\begin{equation}
d_\alpha = \frac{1}{N} \sum_\mathbf{r} S^z_\mathbf{r} S^z_\mathbf{r+\hat \alpha} (-1)^{\mathbf{r}_\alpha}, 
\end{equation}
where $\alpha$ stands for the $x$ or $y$ lattice direction. We scale the temperature in units of $Q$ as $T=c/L$, with $c=2.38$ being
the estimated critical velocity of excitations \cite{Suwa16} (i.e., the system is in the ``cubic'' scaling regime
\cite{Jiang11,Sen15}, as in the case $1/T=L/2$ for the bilayer model in Fig.~\ref{fig:bilayer}).

\begin{figure}
\centering
\includegraphics[width=72mm]{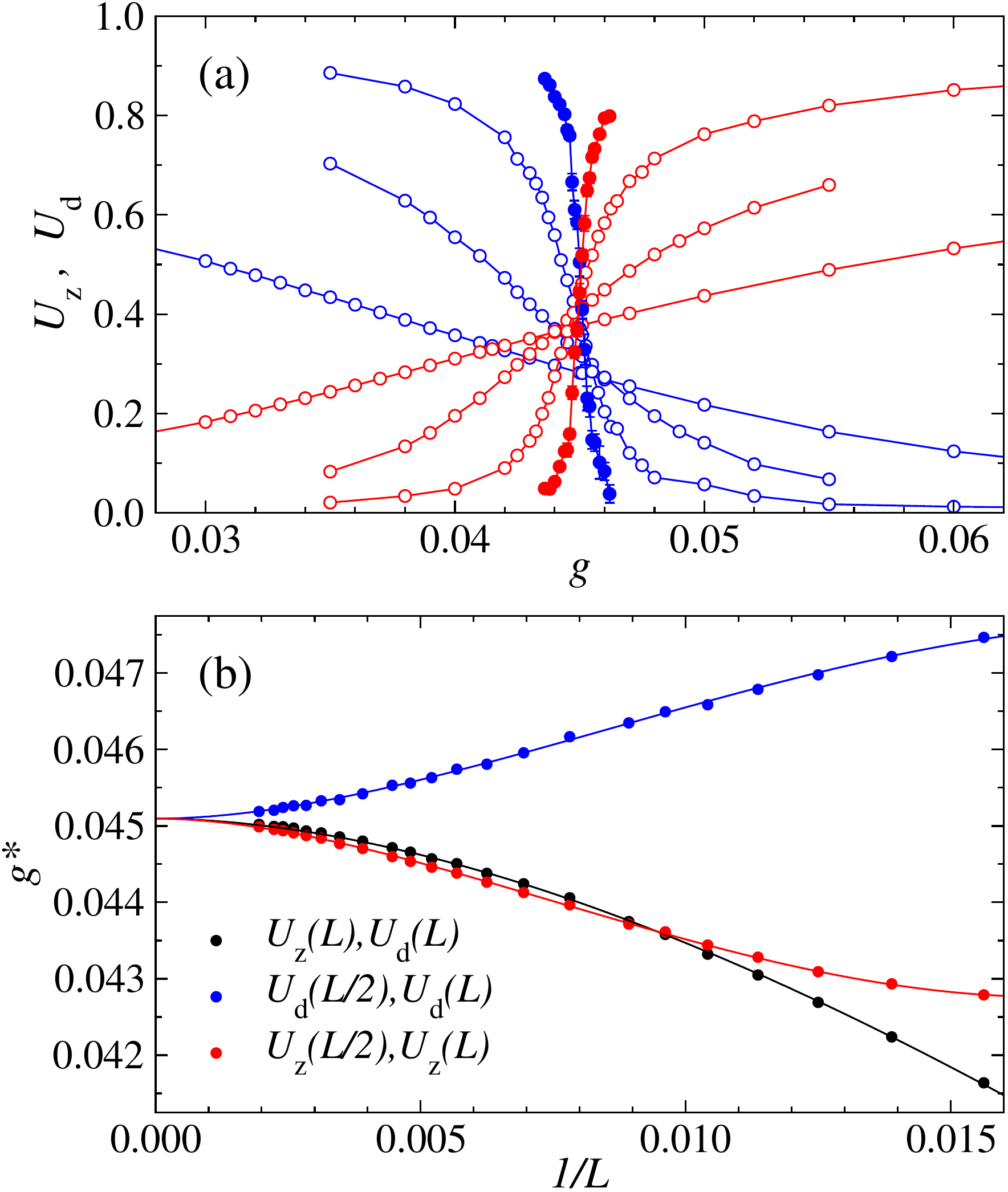}
\caption{(a) Binder cumulants of the AFM (red points) and VBS (blue points) order parameters vs the coupling ratio for system
sizes $L=64$, $128$, $256$, and $512$. The slopes increase with $L$ and the $L=512$ data are shown with solid symbols. (b) Inverse-size dependence
of interpolated crossing points between the two cumulants for given $L$ and for the same cumulant on $L$ and $L/2$ lattices. The curves show 
fits to two power laws for each data set with a common $g_c=g^*(L\to \infty)$ value, resulting in the critical point estimate $g_c =0.04510(2)$.}
\label{fig:gc}
\end{figure}

Early QMC studies placed the VBS--AFM transition at $g_c \approx 0.040$ \cite{Sandvik07,Melko08,Jiang08}, while more recent works show
a somewhat larger value, $g_c \approx 0.045$ \cite{Sandvik10a,Sandvik10b,Suwa16,Shao16}, as a consequences of significant finite-size
corrections. We now have data for system sizes up to $L=512$ and present the Binder cumulants $U_{\rm z}$ and $U_{\rm d}$ defined in the
standard way such that $U_{\rm x} \to 1$ with increasing system sizes if there is order of type x and $U_{\rm x} \to 0$ otherwise;
\begin{equation}
U_{\rm z} = \frac{5}{2} - \frac{5}{6}\frac{\langle m_z^4 \rangle}{\langle m_z^2 \rangle^2},~~~
U_{\rm d} = 2 - \frac{\langle (d_x^2+d_y^2)^2 \rangle}{\langle d_x^2+d_y^2 \rangle^2}.
\end{equation}
Results for several system sizes are shown in Fig.~\ref{fig:gc}(a).

To improve the $g_c$ estimate, we analyze crossing points $g=g^*$ where $U_{\rm z}(g^*,L)=U_{\rm d}(g^*,L)$ and also where (for different $g^*$)
$U_{\rm x}(g^*,L/2)=U_{\rm x}(g^*,L)$ with ${\rm x}={\rm z}$ or ${\rm x}={\rm d}$. As shown in Fig.~\ref{fig:gc}(b), these crossing points flow
to $g_c=0.04510(2)$ as $L \to \infty$. The extrapolation is based on a fit to two power laws for each data set, with a common $g_c$.
Unconstrained fits also result in consistent $g_c$ values. We have excluded small systems until a statistically sound fit is obtained,
with $L \ge 64$ included in the final analysis. From now on we fix the coupling ratio to $g=0.0451 \approx g_c$.

\begin{figure}[t]
\centering
\includegraphics[width=74mm]{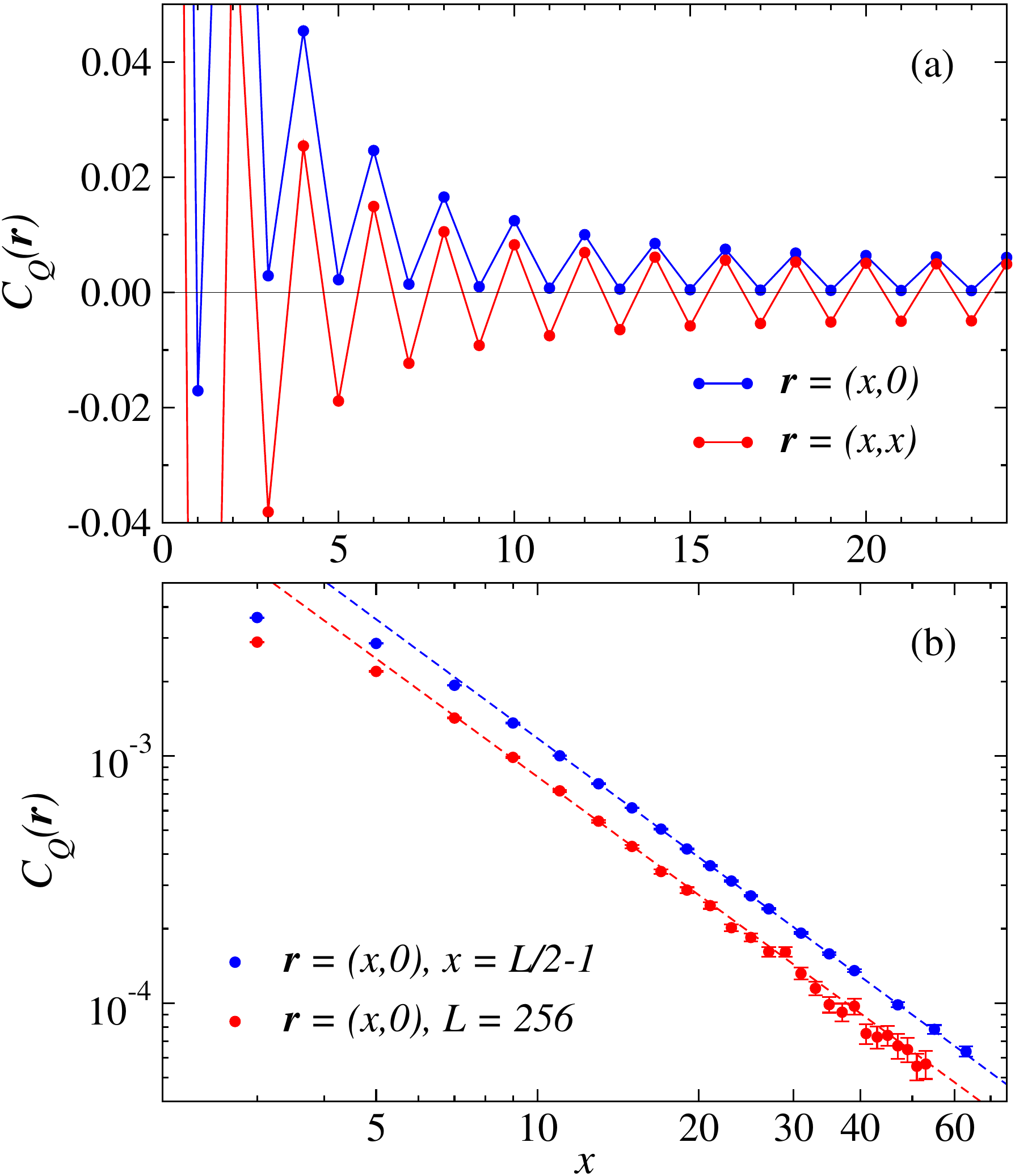}
\vskip-1mm
\caption{Correlation function, Eq.~(\ref{cqdef}), of the $Q$ terms in the critical $J$-$Q$ model ($g=0.0451$).
In (a) results at $\mathbf{r}=(x,0)$ and $(x,x)$ are shown for $L=48$. In (b) results at $\mathbf{r}=(x,0)$ are shown only for
odd values of $x$, with blue points at $x=L/2-1$ for different system sizes $L$ and red points for fixed $L=256$. The lines
in (b) have slope $-2\Delta_Q=-1.60$.}  
\label{fig:x}
\vskip-2mm
\end{figure}

We here examine the correlation function of the $Q$-terms in the Hamiltonian, Eq.~(\ref{jqham}),
\begin{equation}
C_Q(\mathbf{r}_{ij}) = \langle Q_i Q_j\rangle - \langle Q_i \rangle^2,
\label{cqdef}
\end{equation}
which is less noisy than the $J$-energy correlator.
As shown in Fig.~\ref{fig:x}(a), the correlations exhibit strong even-odd oscillations, with amplitude decaying with
the distance. The reason for the oscillating behavior is that the columnar VBS correlations are also detected by the plaquette correlation
function $C_Q(\mathbf{r})$ (for a detailed general discussion of this, see Ref.~\cite{Takahashi20}). In a columnar state with $x$-oriented
dimers, $C_Q(0,y)$ will be small while $C_Q(x,0)$ will have signs $(-1)^x$ due to the dimerization. In an ergodic QMC simulation,
$C_Q(x,y)$ will reflect averaging over states with $x$- and $y$-oriented dimers. The contributions from the VBS order parameter then
cancel in $C_Q(x,0)$ for odd $x$, while $C_Q(x,x)$ retains the VBS contributions with $(-1)^x$ signs.
These behaviors are seen in Fig.~\ref{fig:x}(a), where the amplitude decay is due to the system being a critical VBS.
Since the system has emergent U(1) symmetry of the order parameter \cite{Sandvik07,Jiang08}, we should consider $C_Q(\mathbf{r})$
as averaged over an angle $\phi \in [0,2\pi)$ corresponding to a circular-symmetric distribution $P(d_x,d_y)$. The above mentioned
behaviors of $C_Q(\mathbf{r})$ along the lines $\mathbf{r}=(x,0)$ and $\mathbf{r}=(x,x)$ will hold also in this case.

In addition to the large contributions to $C_Q(\mathbf{r})$ from the VBS order parameter, there should be a uniform component
reflecting the scaling dimension of the full $Q$ operator. Since the VBS contributions are absent at $(x,0)$ with odd $x$, examining the correlations
at these distances is a good way to access the uniform component. In Fig.~\ref{fig:x}(a), small rapidly decaying values are indeed seen,
and in Fig.~\ref{fig:x}(b) the functional form is analyzed on a log-log plot. We use a large system, $L=256$, with $x \ll L$,
as well as $x=L/2-1$ for smaller sizes. In both cases we observe the same algebraic asymptotic decay, and a power-law fit to
the $x=L/2-1$ data for $x>12$ gives $\Delta_Q=0.800(4)$. This scaling dimension corresponds to $1/\nu=2.200(4)$,
in good agreement with the previous (less precise) results for the $J$-$Q$ \cite{Shao16} and loop \cite{Nahum15b} models.

\begin{figure}[t]
\centering
\includegraphics[width=72mm]{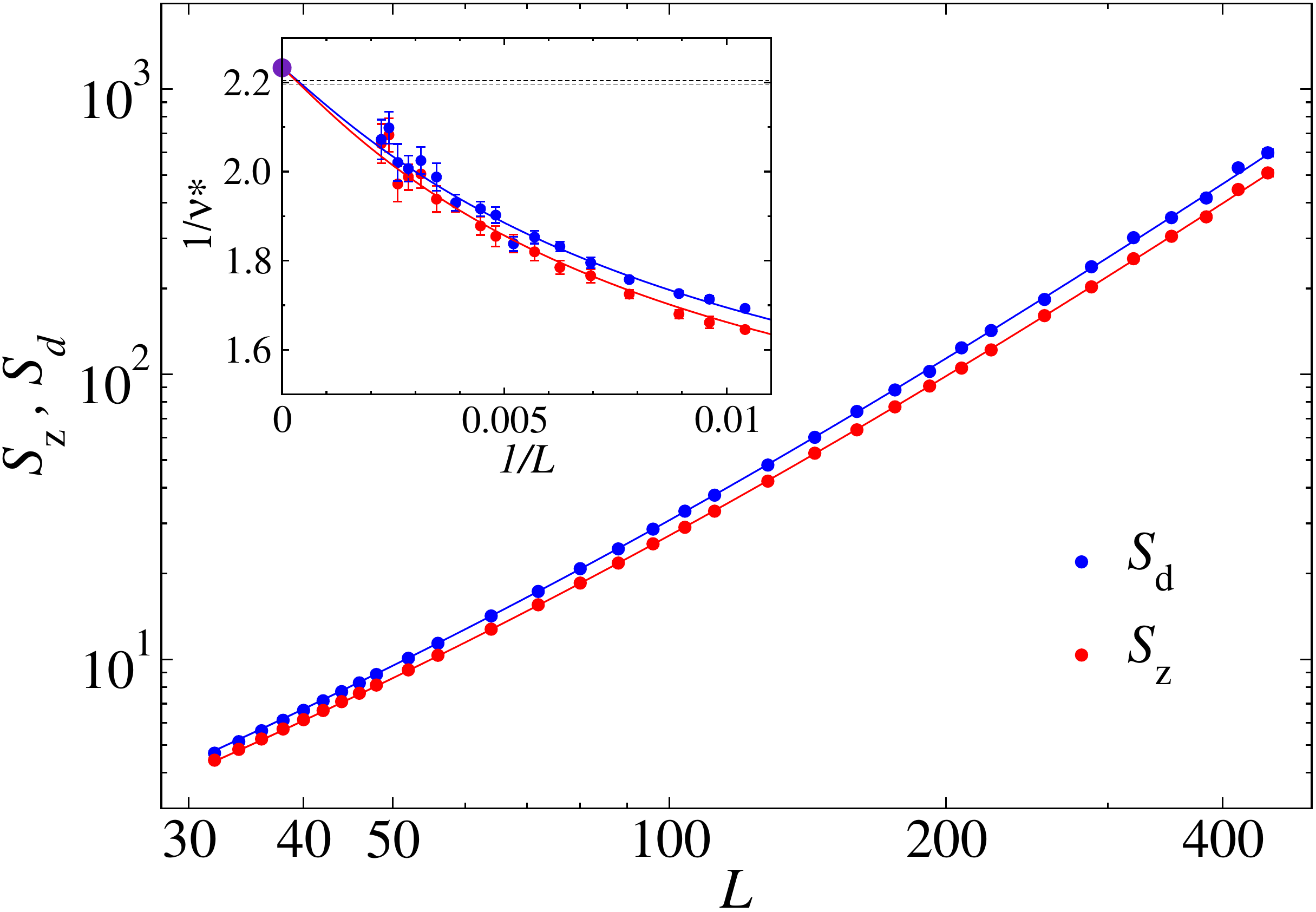}
\vskip-1mm
\caption{Critical cumulant slopes vs the system size. The curves are fits of the $L \ge 64$ data to the
form $aL^{1/\nu}(1 + bL^{-\omega})$, with $1/\nu = 2.23(2)$ (constrained to be the same for both data sets) and $\omega=1.1(1)$ (for both
data sets, not constrained to be the same). The inset shows $1/\nu^* \equiv \ln[S(L)/S(L/2)]\ln^{-1}(2)$ vs $1/L$. The purple circle indicates
the extrapolated exponent $1/\nu=2.23(2)$ and the dashed lines show the values $1/\nu=3-\Delta_Q=2.200 \pm 0.004$ determined in Fig.~\ref{fig:x}.}
\label{fig:y}
\vskip-1mm
\end{figure}

Next we consider the cumulant slopes $S_{\rm x} \equiv dU_{\rm x}/dQ$, ${\rm x} = {\rm d,z}$, computed with direct SSE estimators as
previously done for $S_{\rm z}$ with $L \le 160$ in Ref.~\cite{Shao16}. Here we present results for $L$ up to $L=448$ (our $L=512$
results are too noisy). The slopes should scale asymptotically as $L^{1/\nu}$. In order to account for
the leading correction we also include a second power-law term with smaller exponent, and exclude small systems until good fits are obtained.
The results are shown in Fig.~\ref{fig:y}. The inset shows the same data sets and fits converted into $1/\nu^* \equiv \ln[S(L)/S(L/2)]\ln^{-1}(2)$,
which flows to $1/\nu$ as $L \to \infty$. We note that: (i) $1/\nu = 2.23(2)$ is fully consistent with the previous result from smaller
systems \cite{Shao16}, and (ii) the value also agrees with the above result from the scaling dimension of the $Q$ terms (with deviations
less than 1.5 standard deviations).

While the finite-size corrections in $1/\nu$ obtained from the cumulant slopes in Fig.~\ref{fig:y} are substantial, the corrections to
the $r^{-2\Delta_Q}$ form of the correlation function in Fig.~\ref{fig:x} are very small. The good agreement of the extracted exponents with
the relationship $1/\nu=3-\Delta_Q$ should alleviate any concerns of $1/\nu$ eventually flowing to the value $3$ ($=d$) expected at a conventional
first-order transition (or to $d+1$, as found at an unconventional transition in Ref.~\onlinecite{Zhao19}). Weak
first-order transitions are often most clearly manifested in $1/\nu$ \cite{Iino19}, and the results presented here simply do not
indicate anything unusual. 

Previously, anomalous scaling was found of the order parameters and the spin stiffness
\cite{Jiang08,Sandvik10a,Chen13,Nahum15b}, and it was argued that the standard finite-size scaling hypothesis must be replaced by a form 
taking into account two divergent length scales in a new way \cite{Shao16}. Though this interpretation has not been independently confirmed, the results
presented here reinforce the notion that anomalies are not present in the magnitude $L^{-1/\nu}$ of the critical window. The value $\nu =0.455(2)$ is
still puzzling in the sense that it violates a CFT bound from the bootstrap method \cite{Nakayama16}. This disagreement suggests that the transition
is either not described by a CFT or that the arguments underlying the bound has some loophole. It would be interesting to compute $\nu$ from
the relevant critical correlator for fermionic DQCP candidate models \cite{Liu18}, where the standard finite-size analysis is difficult because
of the rather small accessible lattices. 

\begin{acknowledgments}
{\it Acknowledgments}.---We would like to thank Jun Takahashi for useful discussions.
This work was supported by the NSF under Grant No.~DMR-1710170 and by a Simons Investigator Grant.
The numerical calculations were carried out on the Shared Computing Cluster managed by Boston University's 
Research Computing Services.
\end{acknowledgments}

\end{document}